\documentclass[11pt]{article}
\usepackage{geometry}                % See geometry.pdf to learn the layout options. There are lots.
\usepackage{graphicx}
\usepackage{amssymb}
\usepackage{epstopdf}
\usepackage{epsf}

\title{Spiraling toward market completeness and financial instability}
\author{Matteo Marsili\\
{\em The Abdus Salam International Centre for Theoretical Physics,}\\
{\em Strada Costiera 11, 34014 Trieste, Italy}
}

\begin{document}
\maketitle
%\section{}
%\subsection{}

%"Reuters quoted a financial broker as saying Minerva had become influential enough to deter some investors from buying Korean shares while the government has turned increasingly sensitive to negative reports on the economy, one of the hardest hit in Asia by the global financial crisis." The Korea Times, Nation, Saturday, January 17, 2009

\begin{abstract}
I study the limit of a large random economy, where a set of consumers invests in financial instruments engineered by banks, in order to optimize their future consumption. 
This exercise shows that, even in the ideal case of perfect competition, where full information is available to all market participants, the equilibrium develops a marked vulnerability (or susceptibility) to market imperfections, as markets approach completeness and transaction costs vanish. The decrease in transaction costs arises because financial institutions exploit trading instruments to hedge other instruments. This entails trading volumes in the interbank market which diverge in the limit of complete markets.

These results suggest that the proliferation of financial instruments exacerbates the effects of market imperfections, calling for theories of market as interacting systems. From a different perspective, in order to prevent an escalation of perverse effects, markets may necessitate institutional structures which are more and more conspicuous as their complexity expands. 
\end{abstract}

\newpage

\begin{quotation}
"As products such as futures, options, swaps, and securitized loans become standardized and move from intermediaries to markets, the proliferation of new trading markets in those instruments makes feasible the creation of new custom-designed financial products that improve Òmarket completeness,Ó to hedge their exposures on those products, the producers (typically, financial intermediaries) trade in these new markets and volume expands; increased volume reduces marginal transaction costs and thereby makes possible further implementation of more new products and trading strategies by intermediaries, which in turn leads to still more volume. Success of these trading markets and custom
products encourages investment in creating additional markets and products, and so on it goes,
spiraling toward the theoretically limiting case of zero marginal transactions costs and dynamically
complete markets." \cite{MertonBodie}.
\end{quotation}

The idea that the optimal allocation of goods and resources can be achieved through the expansion of financial markets has deep roots in neo-classical economic theory \cite{Radner,Duffie}. In brief, General Equilibrium Theory (GET) maintains that utility and profit maximization, under perfect competition, ensures optimal outcomes only if all markets exist.
% (put differently, inefficiency, in the GET perspective, arises because some markets do not exist). 
In a non-deterministic world, this entails a quite unrealistic requirement, namely the existence of {\em contingent} commodity markets, i.e. markets, open today, for the delivery of any good at any future date, in any possible state of the world. 

In this framework, financial markets play an important r\^ole, as they {\em substitute} for missing contingent commodity markets, by allowing agents to allocate their wealth to cope with the uncertainty of future spot market prices (see \cite{Radner,Duffie}).
In particular, when the repertoire of trading instrument expands, financial markets approach completeness\footnote{We recall that a complete market is one where any contingent claim can be replicated by a portfolio of traded assets.},
and decentralized utility maximizing behavior achieves GET (Pareto) optimal allocation outcomes.

This theoretical construction sharply contrasts with the 2007-2008 experience, where different forms of {\em market imperfections} have played a major role in turning expanding credit derivative markets into "financial weapons of mass destruction" \cite{Buffet2002}.

Most concerns have focused on the distortions in how financial engineering has been applied, rather than on its theoretical foundations. Several authors have pointed out issues of misaligned incentives, asymmetric information, moral hazard and lack of transparency \cite{GFSR,Rajan2005,Jarrow,failure}.
Such a focus on ``market imperfections'' calls for institutions which can avoid the escalation of perverse effects, leading to systemic crisis. Quoting again from Ref. \cite{MertonBodie}: 
\begin{quotation}
"When particular transaction costs or behavioral patterns produce large departures from the predictions of the ideal frictionless neoclassical equilibrium for a given institutional structure, new institutions tend to develop that partially offset the resulting inefficiencies. In the longer run, after institutional structures have
had time to fully develop, the predictions of the neoclassical model will be approximately
valid for asset prices and resource allocations". 
\end{quotation}
In this perspective, developments in financial engineering are part of the solution, not of the problem\footnote{Quoting from R. J. Schiller ``The history of finance over the centuries has been one of 
gradual expansion of the scope of markets. Over time, 
more and more kinds of risks are traded, and there are 
more and more opportunities for hedging those risks. 
Now is the time to encourage the further development of 
markets in a way that truly democratizes them, that is, so 
that the markets cover the specific risks that ultimately 
matter to individual people."  \cite{Schiller2008}.}
%"[...] %technology will carry us forward into new dimensions of democratized financial sophistication that we cannot now imagine", and that 
%"Now is the time to encourage the further development of markets in a way that truly democratizes them, that is, so that the markets cover the specific risks that ultimately matter to individual people. [...]
%Our society could look forward to nothing less than more stable markets and, in turn, a more rational economy. 
%We would eventually find ourselves forgetting that the kind of massive financial instability infecting our everyday lives is even a potential problem. 
%Modern finance, applied democratically, can relegate these problems to history just as modern medicine, applied widely, has left us forgetting that epidemics of yellow fever and diphtheria ever raged among us." \cite{Schiller2008}.}.
Indeed, in a perfectly competitive market, where full information is available to all market participants, the ``financial innovation spiral'' \cite{MertonBodie} can expand the repertoire of financial instruments, approaching the ideal of a {\em complete} market in which %every risk can be hedged away.
%, because 
any contingent claim can be replicated by a portfolio of traded assets.

%On the theoretical side, \cite{Brock_etal2006} have shown that the introduction of financial instruments in a market with heterogeneous traders, may destabilize the market making it harder for adaptive expectations to converge.

This paper focuses on a theoretical realization of this program, and it analyzes the properties of an economy as financial markets approach completeness. We set ourselves in the ideal -- and seemingly implausible -- long-term scenario where institutions can be designed to guarantee perfect competition, with full information available to all participants. 

The model we focus on is conceptually simple:
%but we analyze it in the limit of highly complex markets, where full heterogeneity of financial instruments is taken into account. This provides a picture at the global scale of the economy, which will allow us to quantify financial stability. 
it describes a single period economy where consumers aim at allocating optimally their wealth in an uncertain future, buying financial instruments which are developed and offered by financial firms -- banks for short -- in a competitive financial sector.
This rather standard setting (see e.g. \cite{Pliska}) is investigated in the complex case of a large random economy, where the number of states and the number of financial instruments both grow very large. In particular, we derive results for ensembles of markets where financial instruments are drawn from a given probability distribution. Put differently, banks develop new instruments through a genuine innovation process, modeled as a random draw of a new asset. 

Taking advantage of statistical physics of disordered systems we are able to characterize the typical properties of the equilibrium in terms of self-averaging quantities (i.e. quantities which satisfy laws of large numbers), and of the relations between them. 

The main result of the paper is that the limit when markets become complete is a singular one. In the space of parameters it separates a stable, arbitrage-free region, from an unstable one. The proliferation of financial instruments brings the market closer to the unstable region, thus eroding systemic stability. 
This entails vanishing transaction costs (risk premium) which arise from the fact that banks can avail of existing assets to hedge the risks of new financial instruments. In a competitive market, the risk premium
charged by banks is proportional to the residual risk, which vanishes as the market becomes complete.

Stability is quantified, in this approach, by the {\em susceptibility} of the equilibrium, i.e. its sensitivity to parameter specification, which diverges as the boundary of the stable and unstable region is approached. 
Such enhanced susceptibility is also what makes it hard, for an adaptive consumer to correctly learn how to invest optimally, when markets are close to being complete. This result is reminiscent of the one discussed in \cite{Brock_etal2006}, who show that the introduction of financial instruments in a market with heterogeneous traders, may destabilize the market making it harder for adaptive expectations to converge\footnote{A similar conclusion is reached in a different class of models of in Ref. \cite{Marsili08SSRN}.}. Indeed, the origin of the instability does not lie in the learning dynamics. It is rather a property of the equilibrium itself.

One particularly catastrophic effect is that, replicating portfolios used by banks to hedge new instruments, require trading volumes, within the financial sector, which diverge as the market approaches completeness\footnote{A different mechanism for a ``lending boom'' in credit derivative markets is discussed in \cite{shin}.}.
Furthermore, the interbank market itself develops a divergent susceptibility, as the theoretical limit of complete markets is approached.

%A similar approach shows that the expansion of derivative markets generates instability and large movements in underlying markets.

In summary, the main point of the paper is that, even in an ideal world, the expansion of financial instruments may bring detrimental ``unintended consequences'' for systemic stability and, in particular, %These results suggest that the proliferation of financial instruments 
it may exacerbate the effects of market imperfections. In order to prevent an escalation of perverse effects, markets may necessitate institutional structures which are more and more conspicuous as they expand. At the same time, the increased sophistication of financial markets calls for theories of financial markets as interacting systems, going beyond the ideal limit of perfect competition.

\section{The model}

In this paper, assets are vehicles for transferring wealth from an initial time $t=0$ (e.g. today) to a future date $t=1$ (e.g. tomorrow) \cite{Pliska}. There are $K$ risky assets whose price $s_k^0$ ($k=1,\ldots,K$) is known at $t=0$. The prices of assets at $t=1$, are instead unknown and are modeled as random variables: We assume the market at $t=1$ can be in one of  $\Omega$ possible states $\omega=1,\ldots,\Omega$, each occurring with probability $\pi^\omega$. Correspondingly, the price of each asset $k$ is $s_k^0+r^\omega_k$ at $t=1$, if state $\omega$ materializes. 
There is also a risk-less asset (bond) which also costs one today and pays one tomorrow, in all states\footnote{This is equivalent to considering, for the sake of simplicity, discounted prices right from the beginning.}. Perfect competition entails that prices are (considered as) exogenous variables, i.e. independent of the behavior of traders. All market participants know the statistics of prices, i.e. they know $r_i^\omega$ for all $i$ and $\omega$ and they know the probability $\pi^\omega$, but they do not know which state will materialize.

\subsection{The financial industry}

Assets are produced -- or engineered -- by financial institutions, either as means to raise investment for firms (equities or stocks) or as means to insure investors against potential risks (e.g. derivatives). For our purpose, the only difference between different types of stocks is the expected return: The expected return of stocks is generally positive $E_\pi[r]>0$, whereas for a financial instruments designed by cover some risk (derivatives), banks will ask for a risk premium, i.e. $E_\pi[r]<0$. 

We assume that banks develop a number $N$ of financial instruments (assets) which are offered to investors. Asset $i=1,\ldots,N$ has a return $r_i^\omega$ in state $\omega=1,\ldots,\Omega$. 
For the sake of simplicity, we are going to take assets with the same risk premium, i.e. we posit that
\begin{equation}
\label{riskpremium}
E_\pi[r_i]=\sum_{\omega=1}^\Omega \pi^\omega r_i^\omega=-\frac{\epsilon}{\Omega}.
\end{equation}
In what follows, returns $r_i^\omega$ are taken as i.i.d. random variables with variance $1/\Omega$, satisfying Eq. (\ref{riskpremium}). We shall come back later to a discussion of the risk premium. For the moment being $\epsilon$ will just be considered as a fixed parameter\footnote{The dependence of the statistics of $r_k^\omega$ on $\Omega$ is designed to have a meaningful limit $\Omega\to \infty$ (see below).}. As we shall see below, not all assets developed by the financial industry will be traded in the market. 

It is worth to remark that innovation in the financial industry is equivalent, in this setting, to a new independent draw from the distribution of assets, which expands the repertoire of available assets $N\to N+1$. The assumption of independence of the $N+1^{\rm st}$ financial instrument from the $N$ existing ones, may be unrealistic. However, as we shall see, whether an innovation is actually adopted (i.e. traded) or not, depends on the whole repertoire of traded assets. Hence, a new {\em successful} financial innovations is not an independent draw. At any rate, our aim is not that of reproducing realistic behavior, but rather that of exploring the consequences of the proliferation of financial instruments in an ideal setting. In this respect, our model definitely captures a genuine discovery process in the financial industry.

When $N$ and $\Omega$ are large, as it will be assumed below, this setting reproduces a situation of great financial complexity.

In this setting, we shall focus on the behavior of the market in terms of two key parameters:
\begin{itemize}
  \item the {\em financial complexity}  $n=N/\Omega$, measured by the number of different financial instruments provided by banks, and
  \item the risk premium $\epsilon$.
\end{itemize}
Competition in the financial industry will likely push the market towards higher values of $n$ and smaller values of $\epsilon$ (see Section \ref{sect_epsilon})

\subsection{Optimizing future consumption}

We imagine there is a population of consumers, described for simplicity by a single representative consumer, whose aim is to maximize the expected utility of consumption $E_\pi[u(c)]$ at $t=1$. Smooth preferences, represented by a monotone increasing, convex utility function $u\in C^2$, are assumed throughout the paper. The quantity $c^\omega$ of consumption good which consumers secure in state $\omega$ depends on the wealth $w_1^\omega$ they have at $t=1$ and on the price $p^\omega$ of the good in state $\omega$: $c^\omega=w_1^\omega/p^\omega$. We assume $p^\omega$ are exogenously given. Specifically, they are drawn independently for each state $\omega$ from a given distribution. The wealth $w_1^\omega$ results from the investment of the initial ($t=0$) wealth $w_0$ of consumers in the financial market. We take for simplicity $w_0=1$ in what follows. Then, if $z_i\ge 0$ is the quantity of asset $i$ bought by consumers,
\begin{equation}
\label{w1}
w_1^\omega=1+\sum_{i=1}^N z_i r_i^\omega.
\end{equation}
The portfolio weights $z_i\ge 0$ are given by the solution of the optimization problem 
\begin{equation}
\label{maxU}
\max_{\vec z\ge 0}E_\pi\left[u\left(\frac{1+\sum_iz_ir_i^\omega}{p^\omega}\right)\right].
\end{equation}
In brief, the consumers exploit the financial market to carry their wealth from $t=0$ to $t=1$ in an optimal manner, in order to cope with the uncertainty of commodity prices\footnote{This set up is rather standard in Asset Pricing Theory \cite{Pliska}. In a General Equilibrium setting, the financial market substitutes missing contingent commodity markets, i.e. markets, cleared at $t=0$, for delivering every commodities at $t=1$ in each state $\omega$. We refer the interested reader to the literature on Radner or sequential equilibria \cite{Radner}.}. 

Inspection of the first order conditions of the problem above reveals two important facts: {\em i)} The consumer will select a subset of $K\le N$ assets which will be traded, whereas all the others will have $z_i=0$. {\em ii)} If a finite solution to the optimization problem (\ref{maxU}) exists, then the subset of $K$ traded assets is arbitrage free, i.e. there exist an Equivalent Martingale Measure (EMM) $q^\omega$ such that $E_q[r_i]=0$ for all assets $i$ which are traded ($z_i>0$). This is because the optimality conditions read
\begin{equation}
\label{optcond}
\frac{\partial}{\partial z_i}E_\pi\left[u\left(c^\omega\right)\right]=\sum_\omega 
\pi^\omega \frac{u'(c^\omega)}{p^\omega}r_i^\omega
~~
\left\{\begin{array}{ccc}=0 & \Leftrightarrow & z_i>0 \\<0 & \Leftrightarrow & z_i=0\end{array}\right.
\end{equation}
which implies that the EMM is given by
\begin{equation}
\label{EMM}
q^\omega=\pi^\omega \frac{u'(c^\omega)}{Q p^\omega},~~~~Q=\sum_\omega 
\pi^\omega \frac{u'(c^\omega)}{p^\omega}.
\end{equation}
Notice that, non traded assets ($z_i=0$) have $E_q[r_i^\omega]<0$, i.e. the price of these instruments ($s_i^0$) is higher than their value $s_i^0+E_q[r_i^\omega]$, according to the EMM, and hence they are not bought by consumers.

\section{Typical behavior of a large complex market}

What are the properties of the market resulting from the interaction between the financial industry and consumers? How, in particular, do these properties change as a function of financial complexity $n=N/\Omega$ and of the risk premium $\epsilon$? 

This question can be answered precisely, in the limit of large random markets $\Omega\to\infty$, because $E[u(c)]$ is expected to satisfy the law of large numbers, i.e. to have a self-averaging behavior. Statistical physics of disordered systems provide the tools to both identify the relevant self-averaging variables -- the so-called order parameters -- as well as the relations coupling them. This approach can be made rigorous following the approach of \cite{Talagrand}. Our interest here is on the behavior of the economy rather than on the mathematics involved, so we shall provide only essential details. The method and the explicit calculation for a problem very similar to (\ref{maxU}) is discussed in detail in \cite{macroecondyn}, to which the interested reader is referred. A short account is given in the appendix.

\subsection{The arbitrage-free region}

Before discussing the general properties of the solution, it is worth to discuss the region where one expects it to exist. Indeed, a solution does not exist if the market is not arbitrage-free, i.e. if investors can buy a portfolio of assets $\vec\zeta=(\zeta_1,\ldots,\zeta_N)$, with $\zeta_i\ge 0$, such that 
\begin{equation}
\label{noarb}
r_\zeta^\omega=\sum_{i=1}^N\zeta_i r_i^\omega\ge 0, ~~~ \forall \omega=1,\ldots,\Omega
\end{equation}
and $r_\zeta^\omega>0$ for at least one state $\omega$. One can compute the volume 
\begin{equation}
\label{Vol}
V(n,\epsilon)=\left |\left\{\vec \zeta\ge 0:~\sum_{i=1}^N\zeta_i r_i^\omega\ge 0~\forall\omega\in\Omega\right\}\right|
%\int_{\vec\zeta\ge 0}d^N\vec\zeta\prod_{i=1}^\Omega\theta\left(\sum_{i=1}^N\zeta_i r_i^\omega\right)
\end{equation}
of portfolios satisfying Eq. (\ref{noarb}) following the same approach as in \cite{macroecondyn}. An account of the necessary steps is given in appendix \ref{appA3}. In brief, one expects that $V(n,\epsilon)=0$ in the arbitrage free regions, whereas outside this region $V(n,\epsilon)\sim D^N$ should shrink as the boundary between the two regions is approached, with $D=|\vec\zeta_1-\vec\zeta_2|$ being the typical distance between two possible portfolios satisfying Eq. (\ref{noarb}). The dependence on $N$ suggests that the quantity with a self-averaging behavior is $\frac{1}{N}\log V(n,\epsilon)$.

\begin{figure}[htbp]
   \centering
   \includegraphics[width=9cm]{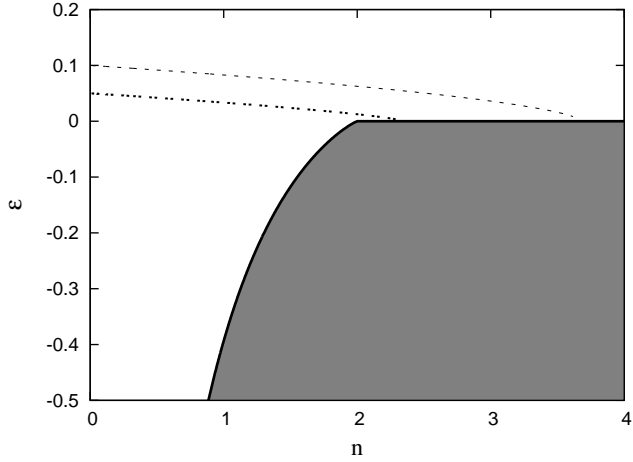} % requires the graphicx package
   \caption{Phase diagram of the economy. The equilibrium is unstable in the shaded region on the bottom right. The dashed and dotted lines correspond to ``trajectories'', where the risk premium $\epsilon$ is determined endogenously, as discussed in Section \ref{sect_epsilon}, with $\gamma=0.05$ (bottom) and $\gamma=0.1$ (top).
   }
   \label{phasediag}
\end{figure}

The analysis of this quantity, averaged over the ensemble of realizations of the economy (i.e. on $r^\omega_i$ and $p^\omega$) leads to the boundary of the arbitrage free region, which is shown in Fig. \ref{phasediag}.

For $\epsilon>0$ the market is always arbitrage free, as banks demand a positive risk premium for all assets. When $\epsilon<0$, instead, as the number of assets introduced by banks increases, the market approaches a state where an arbitrage appears with probability one. The properties of the economy are radically different when the boundary is approached for $\epsilon<0$ or for $\epsilon>0$.

\subsection{Approaching complete markets}

A similar analysis can be applied to Eq. (\ref{maxU}), as discussed in the appendix. 
Let us focus our discussion on the behavior of the following quantities 
\begin{equation}
\label{Rsigmaphi}
R=\frac{\epsilon}{\Omega}\sum_{i=1}^N z_i, ~~~~\sigma^2_q=\Omega \sum_{\omega=1}^{\Omega}\left(q^\omega-\pi^\omega\right)^2,~~~~\phi=\frac{|\{i:~z_i>0\}|}{\Omega}
%,~~~~G=\frac{1}{N}\sum_{i=1}^N z_i^2
\end{equation}
which are respectively the total revenue $R$ of the financial sector, the deviation of the EMM from the empirical measure and the degree of market completeness. Note, in particular, that $\phi=1$ corresponds to a situation where the number of traded assets equals the number of states, i.e. to complete markets. 

In addition, one quantity which turns out to play a key role is the {\em susceptibility}. In order to define this quantity, one first introduces a small variation in the utility $E[u(c)]\to E[u(c)]+\vec h\cdot \vec z$. Then one computes the variation of the solution $\vec z(\vec h)$ of the corresponding optimization problem
\begin{equation}
\label{chidef}
\chi=\left.\frac{1}{N}\sum_{i=1}^N\frac{\partial z_i}{\partial h_i}\right|_{\vec h=0}
\end{equation}
for vanishing $\vec h$. $\chi$ measures the sensitivity of the solution $\vec z$ to the parameters which specify the original problem\footnote{$\chi$ captures sensitivity wrt the definition of the utility function. Likewise, it is possible to define the sensitivity wrt to other parameters. For instance, the variation of the portfolios $\vec z$ for a small change $r_i^\omega\to r_i^\omega+\delta r_i$ of returns, corresponds to a change $c^\omega\to c^\omega+\delta\vec r\cdot\vec z/p^\omega$ in consumption and, to leading order, to a change
\[
E[u(c+\delta c)]=E[u(c)]+E[u'(c)/p]\delta\vec r\cdot\vec z+O(\delta\vec r^2)
\]
in the utility. 
Hence the sensitivity to the specification in returns is given by
\[
\left.\frac{1}{N}\sum_{i=1}^N\frac{\partial z_i}{\partial \delta r_i}\right|_{\delta \vec r=0}=E[u'(c)/p]\chi.
\]
}. 

The analysis of the optimization problem, discussed in appendices \ref{appA2} and \ref{appA22}, allows us to state the following:

\begin{quotation}
The locus on which markets become complete coincides with the line $\epsilon=0$, $n\ge 2$ which separates the no-arbitrage region from the unstable one. In the limit of complete markets, i.e. as $\phi\to 1^-$, 
\begin{enumerate}
\item The equilibrium becomes more and more fragile, in the sense that the susceptibility $\chi$ diverges as $\phi\to 1$.
\item Consumers exhibit a risk neutral behavior, in the sense that $q^\omega\to\pi^\omega$ (i.e. $\sigma\to 0$)
\end{enumerate}
\end{quotation}

The convergence of the EMM $q$ to the empirical measure $\pi$ implies, by Eq. (\ref{EMM}) that the marginal utility of consumption in each state is proportional to the price of commodity in that state, with a state independent proportionality constant. 

\begin{figure}[htbp]
   \centering
   \includegraphics[width=9cm]{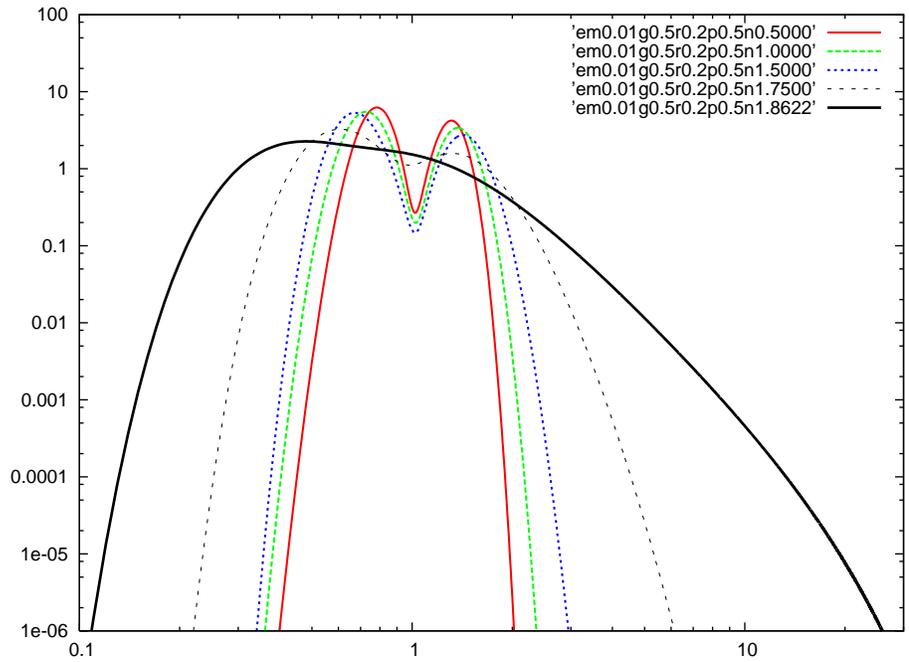} % requires the graphicx package
   \caption{Probability density of consumption $c^\omega$ for $\epsilon=-0.01$ for $n=0.5,1.0,1.5,1.75$ and $1.86$. The probability distribution broadens as $n$ increases approaching the value $n_c\cong 1.92$ where arbitrage opportunities appear. The distribution refers to the case discussed in the text:
   $\pi^\omega=1/\Omega$,  $u(c)=2(\sqrt{c}-1)$, with a bimodal distribution of prices, i.e. $p^\omega=1\pm \varrho$ with probability $1/2$ and $\varrho=0.2$.}
   \label{figc}
\end{figure}

The susceptibility $\chi$ diverges also upon approaching the unstable region with $\epsilon<0$, but the market is not complete (i.e. $\phi<1$) neither $\sigma$ vanishes, on the boundary of the unstable region for $\epsilon<0$. The reason for the divergence of $\chi$ for $\epsilon<0$, is that arbitrage opportunities arise, corresponding to portfolios $\vec z$ with divergent weights, which deliver boundless consumption in some states. The evolution of the probability distribution of final consumption as the economy approaches the unstable region for $\epsilon<0$, is shown in Fig. \ref{figc}.

Specific results for any particular choice of the probability distribution $\pi^\omega$, the utility function $u(c)$ and of the distribution of prices $p^{\omega}$ can be derived solving Eqs. (\ref{eqlam} -- \ref{eqkappa}) given in appendix \ref{appA2}. Here we confine our discussion to the representative case of $\pi^\omega=1/|\Omega|$,  constant relative risk aversion utility $u(c)=(c^{\gamma}-1)/\gamma$ with $\gamma=1/2$ and bimodal distribution of prices, i.e. $p^\omega=1\pm \varrho$ with probability $1/2$\footnote{It might be interesting to note that in the limit $\gamma\to 0$ where $u(c)\to \log c$, the portfolio $\vec z\to 0$ vanish for all $\epsilon>0$. This can be verified from the first order conditions of the original problem Eq. (\ref{maxU}).}. 

\begin{figure}[htbp]
   \centering
   \includegraphics[width=9cm]{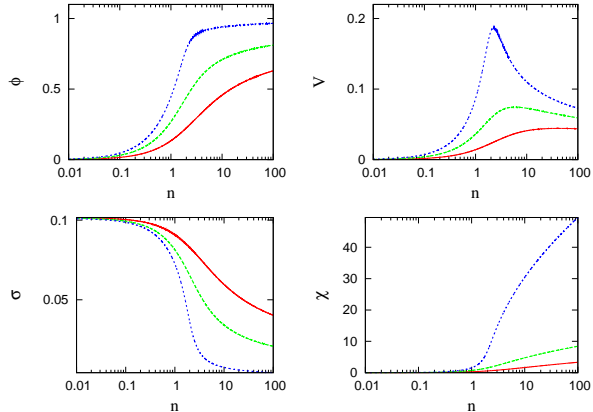} % requires the graphicx package
   \caption{Behavior of the economy as a function of $n$, for different values of $\epsilon>0$: market completeness $\phi$ (top - left), total volume $V=R/\epsilon$ (top - right), distance between the EMM and the empirical measure $\sigma$ (bottom - left) and susceptibility $\chi$ (bottom - right). Curves refer to $\epsilon=0.01,0.05$ and $0.1$ (from bottom to top for $\sigma$ and from top to bottom for the other plots).}
   \label{figfixedeps}
\end{figure}

Fig. \ref{figfixedeps} shows the behavior of $R$, $\phi$, $\sigma$ and $\chi$ for different values of $\epsilon>0$, as a function of $n$. This illustrates how the equilibrium changes when the repertoire of financial instruments expands. 
Even though our picture refers to equilibrium results, it is tempting to interpret this in terms of 
%an {\em adiabatic} 
slow dynamics induced by technological innovation in the finance sector \footnote{Slow here means that we assume that the economy converges to the equilibrium for each $N$ in a time which is much shorter than that for the introduction of the $N+1^{\rm st}$ financial instrument.}, the number of traded assets (i.e. $\phi$) increases and it saturates to a value which approaches the number of states $\Omega$ as $\epsilon\to 0$. Correspondingly, the EMM draws closer to the empirical measure $\pi$ (i.e. $\sigma\searrow n$). This is accompanied by a rise in the susceptibility $\chi$, which gets steeper and steeper as $\epsilon$ decreases. Interestingly, the total volume of assets bought by consumers, and hence the expected revenue $R$ of banks, has a non-monotonic behavior, which is more and more pronounced as $\epsilon\to 0$. This suggests that, for a fixed risk premium $\epsilon$, innovation in the financial sector is profitable only when $n$ is not too large. The increase in the average trading volume of an existing trading instrument, when a new instrument is introduce, signals the presence of non-trivial complementarity effects, when $n$ is small enough. On the other hand, for larger values of $n$, the expansion of the repertoire of instruments brings about a reduction in the volume $z_i$ of trading in already existing assets, an indication of saturation of the financial market.

\section{Risk premium and competitive prices of assets}
\label{sect_epsilon}

Indeed, the risk premium $\epsilon$ is not an independent variable, but it rather depends on the stock of instruments which banks have at their disposal. There are several possible arguments supporting this view. Here we focus on a very simple mechanism, which arises from the fact that banks themselves can buy assets to hedge the risk of new financial instruments they introduce. In other words, each unit of a new asset $r^\omega_{N+1}$ can be backed by a portfolio $\vec w$ of existing assets, which banks can optimize to reduce the risk. It is reasonable to assume that the portfolio is composed of traded assets only, i.e. those with $z_i>0$, the other ones being illiquid. Hence banks have $K=\Omega \phi$ assets at their disposal and the larger is the stock of existing traded assets the smaller will the residual risk associated with the new trading instrument be. Indeed this argument goes for all assets and it ultimately implies that, if the risk premium $\epsilon$ is proportional to non-diversifyable risk, then $\epsilon$ should decrease with $n$. In other words, when $n$ increases, the economy should follow a trajectory in the $(n,\epsilon)$ plane of Fig. \ref{phasediag}, with $\epsilon\searrow n$. This is consistent with the expectation of markets ``spiraling toward the theoretically limiting case of zero marginal transactions costs and 
dynamically complete markets" \cite{MertonBodie}. The results of the previous section, however, imply that when markets approach completeness their stability gets eroded at the same time. Hence the path to complete markets leads indeed to financial instability.

Though this expectation is rather general, the present framework allows us to provide a concrete realization of this mechanism. Let us imagine a financial sector where banks value financial instruments on the basis of a simple mean variance utility function
\begin{equation}
\label{aaa}
\upsilon(\theta)=E[\theta]-\frac{\gamma}{2}E\left[(\theta-E[\theta])^2\right]
\end{equation}
for any contingent claim $\theta^\omega$ and with $\gamma$ the (absolute) risk aversion coefficient.
When issuing the $N+1^{\rm st}$ instrument, banks need to evaluate a contingent claim
\begin{equation}
\label{theta}
\theta^\omega_w=-r_{N+1}^\omega+\sum_{i:z_i>0}w_i r_i^\omega
\end{equation}
where $r_{N+1}^\omega$ is the return which the bank should deliver to consumers, for each unit of the $N+1^{\rm st}$ instrument which they sell, and the second term is what the bank will receive from the portfolio $\vec w$ of assets it has built in order to hedge the new instrument. Since $E[r_i]=E[r_{N+1}]=-\epsilon/\Omega$, the expected utility is
\begin{equation}
\label{meanvar}
\upsilon(\vec w)=\frac{\epsilon}{\Omega}\left(1-\sum_{i:z_i>0}w_i \right)-
\frac{\gamma}{2}E\left[(\theta_w-E[\theta_w])^2\right].
\end{equation}

We restrict our attention to portfolios $\vec w$ with $\sum_i w_i=0$. This implies that, in order to hedge the new instrument, the bank will need to buy and sell other assets from and to other banks, but that the net trading at $t=0$ is zero\footnote{Alternatively, one can think of $\vec w$ as being a replicating portfolio used only for (level 1) evaluation and not really traded.}. Hence $\vec w$ represents trading in the interbank market.

Hence, banks will find the portfolio $\vec w$ which minimizes the residual risk $\Sigma^2(\vec w)=E\left[(\theta_w-E[\theta_w])^2\right]$. Competition in the financial sector will then drive the risk premium to the limit of vanishing banks' profits, i.e. $\min_{\vec w}\upsilon(\theta_w)=0$. This implies that
\begin{equation}
\label{mineps}
\frac{\epsilon}{\Omega}=\gamma \min_{\vec w}\Sigma^2(\vec w).
\end{equation}

The problem in Eq. (\ref{mineps}) can again be tackled with the same techniques we used to derive the results in previous sections. We refer the interested reader to appendix \ref{appA3}, and state the results:

\begin{quotation}
%{\em Proposition 2:} 
In a competitive financial industry, where banks use mean-variance profit functions, the risk premium which is charged to consumers is 
\begin{equation}
\label{epssig}
\epsilon=\frac{\gamma}{2}(1-\phi)
\end{equation}
and it vanishes in the limit of complete markets.

As the market becomes complete, both the volume of trading in the interbank market 
and the interbank susceptibility $\chi_w$ diverge. 
\end{quotation}

\begin{figure}[htbp]
   \centering
   \includegraphics[width=10cm]{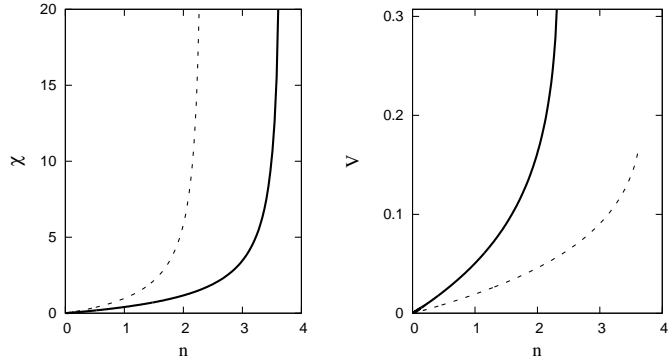} % requires the graphicx package
   \caption{Susceptibility $\chi$ (left) and volume $V=R/\epsilon$ (right) of consumers' portfolios (Eqs. \ref{Rsigmaphi}, \ref{chidef}) as a function of $n$, along the two trajectories depicted in Fig. \ref{phasediag}. The risk premium $\epsilon$ is determined as in Eq. (\ref{mineps}) with $\gamma=0.05$ (dashed line) and $0.1$ (full line). The susceptibility $\chi_w$ and the weights of hedging portfolios diverge, as $\phi\to 1$, according to Eq. (\ref{gwchiw}).}
   \label{figdyneps}
\end{figure}

The first part of the statement reflects the expectation that transaction costs should vanish as the theoretical limit of complete markets is approached. The second part indicates that, in this limit, the volume of trading in the interbank market diverges. More precisely, one finds that
\begin{equation}
\label{gwchiw}
\sum_i w_i^2=\frac{\phi}{1-\phi},~~~~\chi_w=\left.\frac{\delta w}{\delta h}\right|_{h=0}=
\frac{\phi}{\gamma(1-\phi)}
\end{equation}
where the definition of $\chi_w$ is analogous of that given in Eq. (\ref{chidef}) for $\chi$.
The volume implied by hedging each new financial instrument diverges as the market becomes complete\footnote{Alternatively, one may assume that banks maximize $\upsilon(\vec w)$ without the constraint $\sum_iw_i=0$. This leads to very similar results, with $\epsilon=\frac{\gamma}{2\Omega}(1-\phi)$ smaller by a factor $\Omega$. Furthermore, at the optimum one finds $\sum_iw_i<0$. This means that the strategy which maximizes banks expected profit entails a net sale of  ``old'' financial instruments, for each unit of the new one. This excess supply, however, has no counterpart that can absorb it, in the present model.}.

Two trajectories of the economy in the $(n,\epsilon)$ plane, derived from Eq. (\ref{epssig}) for risk aversion coefficients $\gamma=0.05$ and $0.1$, are shown in Fig. \ref{phasediag}. Fig. \ref{figdyneps} reports the value of $\chi$ and $V$ along these lines. All these trajectories ultimately terminate on the line of complete markets ($\phi\to 1$), which separates the no-arbitrage from the unstable region. As this line is approached the portfolio $\vec z$ of consumers becomes unstable, but the volume $V=R/\epsilon$ of consumers portfolios (Eq. \ref{Rsigmaphi}) remains finite. 
Hedging and pricing in the financial industry becomes even more problematic. Not only the susceptibility $\chi_w$, but also the weights $w_i$ of hedging portfolios diverge as $\phi\to 1$, as specified by Eq. (\ref{gwchiw}). 
Each new financial instrument engenders trading volumes, on other assets, which diverge as the market becomes complete. 

\section{Conclusion}

The main result of this paper is that, even in the best of all possible worlds -- i.e. in a regime of perfect competition, where full information is available to all market participants -- the proliferation of financial instruments potentially leads to instability at the system level. Instability is quantified in terms of the susceptibility, i.e. the response of the equilibrium to a perturbation or mis-specification in the underlying parameters of the model. As the market approaches completeness, the susceptibility diverges.

The reason why the divergence of the susceptibility $\chi$ is at the root of financial instability goes along the following argument. 
The mathematical model we are discussing is a rather idealized picture of reality. Still, one may think it is a zeroth order approximation of real financial markets, because different parties can learn or figure out what the equilibrium should be. However, when the susceptibility $\chi$ is large, a small inaccuracy in the estimate of statistical parameters of prices may result in a sizeable deviation of the portfolio $\vec z$ from the optimal one. 
In the case when the market interaction is repeated many periods, investors may fail to correctly learn how to invest optimally if the susceptibility is large. This is because statistical errors, which are unavoidable in practical situations, will be amplified by their learning behavior if $\chi$ is large enough. This, arguably, is the origin of the results in Ref. \cite{Brock_etal2006}: The onset of the large fluctuations is not related to the particular adaptive learning scheme assumed, but rather to the nature of the equilibrium which agents are trying to learn. 

Second, we find that as the repertoire of financial instruments expands, the risk premium decreases, as suggested in Merton and Bodie \cite{MertonBodie}. The equilibrium then approaches the theoretical limit of complete markets, which is plagued by the instability of the portfolio of consumers. 
In the specific implementation of this picture which we have presented, the decrease of risk premia when the number of traded assets increases, entails that replicating portfolios which are used to hedge new instruments, have weights whose absolute size with $n$. This requires trading volumes among banks in the financial industry which diverge as the market approaches completeness. Such interbank market itself becomes less and less stable (i.e. with a divergent susceptibility) as the theoretical limit of complete markets is approached. It's tempting to put this result in relation with the explosion of trading in credit derivatives among financial institutions, which we have witnessed in recent years (see e.g. \cite{GFSR}).

The extension of this approach to more interesting and realistic cases of not perfectly liquid market and asymmetric information is an interesting avenue of future research. 
At a qualitative level, both these effects entail market imperfections which would displace the equilibrium from the one with perfect competition and full information. Within a na\"ive {\em linear response} theory, it is reasonable to assume that, given the results discussed here, the effects of market imperfections will be more and more severe the closer the economy is to one with complete markets. This suggests that the proliferation of financial instruments may exacerbate the effects of market imperfections, or call for tighter regulations in order to prevent an escalation in their effects.

\appendix

\section{The statistical mechanics approach}
\label{app0}

The typical properties of solutions of Eq. (\ref{maxU}), where $r_i^\omega$ are i.i.d. random variables with variance $1/\Omega$, satisfying the constraint 
Eq. (\ref{riskpremium}) for all $i$, is derived from the study of the {\em partition function}
\begin{equation}
\label{Zbx}
Z(\beta,x_{0})=
\int_{\vec z\ge 0}\! d\vec{z}\int_{\vec x\ge 0}\! d\vec{x}
\prod_\omega e^{\beta u(x^\omega/p^\omega)}\delta\left(x^\omega-x_{0}-\sum_{i=1}^N z_i r_i^\omega\right).
\end{equation}
This defines a Gibbs measure 
\begin{equation}
\label{Gibbs}
P(\vec z)=\frac{1}{Z(\beta,x_{0})}e^{\beta \sum_{\omega}u\left(\left(x^\omega-x_{0}-\sum_{i=1}^N z_i r_i^\omega\right)/p^{\omega}\right)}
\end{equation}
over portfolios $\vec z$. 
For large $\beta$ this is dominated by portfolios $\vec z$ for which the utility is maximal. Therefore
\begin{equation}
\label{app0eq1}
\max_{\vec z\ge 0}E_{\pi}\left[u\left(\frac{1+\sum_{i}z_{i}r_{i}^{\omega}}{p^{\omega}}\right)\right]=\lim_{\beta\to\infty}\frac{1}{\beta\Omega}\log Z(\beta,x_{0}=1).
\end{equation}
The partition function also provides information on the volume $V(n,\epsilon)$ of portfolios $z_i$ which violate the no-arbitrage condition. Indeed with $\beta=x_{0}=0$ we have 
\[
Z(\beta=0,x_{0}=0)=\int_{\vec z\ge 0}\! d\vec{z}\prod_{\omega=1}^{\Omega}
\theta\left(\sum_{i=1}^N z_i r_i^\omega\right)=V(n,\epsilon)
\]
with $\theta(x)=\int_0^{\infty}\!dx'\delta (x-x')$ being one if $x\ge 0$ and zero otherwise. Therefore, the knowledge of the statistical properties of $Z(\beta,x_{0})$ provides us with a characterization of the solution of both problems\footnote{Note that, strictly speaking, the limit $\beta\to 0$ of $Z(\beta,0)$ diverges, as the integrals are unbounded. This divergence is easily cured by introducing a cutoff $z_i\le Z$ in the integrals, and then letting $Z\to\infty$.}.

$E_\pi[u(c)]=\frac{1}{\Omega}\sum_\omega u(c^\omega)$ is a random variable, which depends on the draw of the particular realization $\hat r\equiv\{r_{i}^{\omega}\}$ of the returns of instruments and of commodity prices $\vec p=\{p^\omega\}$. In this respect, $u(c^\omega)$ can be regarded as weakly dependent random variables (the only dependence arises from Eq. (\ref{riskpremium}), with first and second moment. Hence, $E_\pi[u(c)]$ satisfies the law of large numbers, i.e. it converges, as $\Omega\to\infty$ to a constant, which is almost surely independent of the specific realization of market returns $\hat r$ and commodity prices $\vec p$. This is reflected in the fact that the explicit calculation of $\log Z$ in effect depends on quantities which satisfy laws of large numbers, and depend only on the first two moments of the distribution of returns $r_{i}^{\omega}$. 

The explicit calculation of the expected value of $\log Z$ over the realization of the returns and prices follows precisely the same steps as those outlined in detail in \cite{macroecondyn}, and will not be repeated here. 

\subsection{The optimal investment problem}
\label{appA2}

The final result, for $x_{0}=1$ and $\beta\to\infty$, is given by the solution of the reduced problem
\begin{equation}
\lim_{N\to\infty} \left\langle E\left[u\left(\frac{1+\sum_{i=1}^Nz_i r_i^\omega}{p^\omega}\right)\right]\right\rangle_{\hat r,\vec p} =  \max_{G,\kappa,\lambda,\sigma,\chi,\hat\chi}\psi(G,\kappa,\lambda,\sigma,\chi,\nu) \label{maxUpsi}
\end{equation}
where the average is taken over the realization of instruments $\hat r$ and commodity prices $\vec p$ and the function $\psi$ is given by
\begin{eqnarray}
\psi & = & n\left\langle\max_{z\ge 0}\left[(t\sigma-\epsilon\lambda)z-\frac{\nu}{2}z^2\right]\right\rangle_t+\frac{n}{2}G\nu+\kappa\lambda-\frac{\chi\sigma^2}{2}-\frac{1}{2}\chi \lambda^2 \label{h}\\
 & + & \left\langle\max_{c\ge 0}\left[u(c)-\frac{(cp-1+\kappa+\sqrt{n G}t)^2}{2\chi}\right]\right\rangle_{t,p}.\nonumber
\end{eqnarray}
Here $\langle\ldots\rangle_{t}$ stands for expected values taken over a Gaussian random variable $t$, with mean zero and unit variance, and, in the last term, 
$\langle\ldots\rangle_{t,p}$ contains an additional expectation over the distribution of prices $p=p^\omega$. 

Before discussing the meaning of the different parameters entering into $\psi$, it is worth to notice that the variables $z$ and $c$ which appear in the maximization problems of Eq. (\ref{h}) are indeed the variables $z_{i}$ and $c^{\omega}=x^{\omega}/p^{\omega}$ which appear in Eq. (\ref{Zbx}). Therefore, the distribution of $t$  induces a distribution $P\{z_{i}\in [z,z+dz)\}  =  \rho_{z}(z)dz$ in the portfolios $z_i$, with  
\begin{eqnarray}
\rho_{z}(z) & = & \left\langle\delta\left(z-z^{*}(t)\right)\right\rangle_{t} \\
z^{*}(t) & = & \frac{\sigma t -\epsilon\lambda}{\nu}\theta(\sigma t-\epsilon\lambda).
\label{FOCz}
\end{eqnarray}
Likewise, the distribution $P\{c^{\omega}\in [c,c+dc)\}  =  \rho_{c}(c)dc$  of consumption levels $c^{\omega}$, is derived from that of $t$ and $p$, as
\begin{equation}
\rho_{c}(c)  =  \left\langle\delta\left(c-c^{*}(t,p)\right)\right\rangle_{t,p} 
\end{equation}
where $c^{*}(t,p)$ is the solution of the first order conditions 
\begin{equation}
\label{FOCc}
\chi\frac{u'(c)}{p}=cp-1+\kappa+\sqrt{nG}t
\end{equation}

In brief, the statistical mechanics approach transforms an optimization problem in $N$ variables, when $N\to\infty$, into the maximization of a function over a finite number of variables, six in our case. The meaning of these variables can be elucidated by inspection of the first order conditiond for the maximization of $\psi$ (the saddle point equations), which are given by
%\begin{eqnarray}
%\lambda & = & \langle u'(c^*)/p\rangle_{t,p} \label{eqlam}\\
%\nu & = & \frac{1}{\sqrt{nG}} \langle u'(c^*)t/p\rangle_{t,p}=
%\left\langle\frac{u"(c^*)}{\chi u"(c^*)-p^2}\right\rangle_{p,t}
%\label{eqnu}
%\\
%\sigma^2 & = & \left\langle[u'(c^*)/p]^2\right\rangle-\left\langle u'(c^*)/p\right\rangle^2\\
%G & =& \langle {z^*}^2\rangle_t\label{eqG}\\
%\chi & = & \frac{n}{\sigma}\langle z^* t\rangle_t=
%\frac{n}{\nu}{\rm Prob}\{z^*>0\}
%\label{eqchi}\\
%\kappa & = & \lambda\chi+n\epsilon\langle z^*\rangle_t. \label{eqkappa}
%\end{eqnarray}
%In Eqs. (\ref{eqnu},\ref{eqchi}) we have used the property $\langle f(t)t\rangle_{t}=\left\langle\frac{\partial f}{\partial t}\right\rangle_{t}$ of gaussian integrals, which is easily shown by integration by parts, and we have used the expressions for $z^{*}$ and $c^{*}$ given above. 
\begin{eqnarray}
\lambda & = & \langle u'(c^*)/p\rangle_{t,p} \label{eqlam}\\
\nu & = & \frac{1}{\sqrt{nG}} \langle u'(c^*)t/p\rangle_{t,p}
\label{eqnu}
\\
\sigma^2 & = & \left\langle[u'(c^*)/p]^2\right\rangle-\left\langle u'(c^*)/p\right\rangle^2\\
G & =& \langle {z^*}^2\rangle_t\label{eqG}\\
\chi & = & \frac{n}{\sigma}\langle z^* t\rangle_t\label{eqchi}\\
\kappa & = & \lambda\chi+n\epsilon\langle z^*\rangle_t. \label{eqkappa}
\end{eqnarray}

The function $\psi$ can be computed numerically to any degree of accuracy, and the maximization problem in the r.h.s. of Eq. (\ref{maxUpsi}) can be carried out numerically. Still it is possible to derive a great deal of insight from inspection of the structure of the maximization problem.

\subsubsection{Interpretation of the parameters and the susceptibility}
\label{appchi}

For what we have said above, the parameters $G$ and $\sigma/\lambda$ have a direct interpretation in terms of the original problem. $G$ is introduced, in the calculation imposing the constraint
\begin{equation}
\label{Gdef}
G=\frac{1}{N}\sum_{i=1}^{N}z_i^{2}
\end{equation}
and indeed Eq. (\ref{eqG}) equates $G$ to the average of the squares of the portfolio weights. Likewise, $\sigma/\lambda$ is the standard deviation of the equivalent martingale measure across states, which is also a measure of the distance of $q^{\omega}$ from the real measure $\pi^{\omega}=1/\Omega$. 

A similar interpretation is possible for the parameter $\chi$. This enters in the calculation of $Z(\beta,1)$, as
\begin{equation}
\label{chidef1}
\chi=\frac{\beta}{2N}\sum_{i=1}^{N}\left(z_{i}-z_{i}'\right)^{2}
\end{equation}
where $\vec z$ and $\vec z'$ are two samples from the Gibbs measure Eq. (\ref{Gibbs}). As $\beta\to\infty$ the two solutions converge to the same one, with their squared distance vanishing as $1/\beta$. There is a more intuitive interpretation of $\chi$ as a {\em response function} or {\em susceptibility}:

Consider the variation of the portfolio weights $\vec z$, under the change $u(c^{\omega})\to u(c^{\omega})+\sum_{{i=1}}^{N}h_i z_{i}$ in the utility\footnote{For small $\vec h$, this corresponds to a change in the perception of the utility of instrument $i=1,\ldots,h_{N}$ or, to leading order in $\vec h$, to a change in the returns $r_{i}^{\omega}\to r^{\omega}_{i}+h_{i}/E_{\pi}[u'(c^{\omega})/p^{\omega}]+O(\vec h^{2})$.}. Let $Z(\beta,x_{0},\vec h)$ be the modified partition function, which is obtained introducing an additional factor $e^{\beta \vec h\cdot \vec z}$ inside the integral of Eq. (\ref{Zbx}). The value of the portfolio weight $z_{i}$ can now be computed as follows
\begin{equation}
\label{eqzi}
z_{i}=\lim_{\beta\to\infty}\left.\frac{1}{\beta}\frac{\partial}{\partial h_{i}}\log Z(\beta,1,\vec h)\right|_{\vec h=0}
=\lim_{\beta\to\infty}\left.\frac{1}{Z(\beta,1,\vec h)}\int_{\vec z\ge 0}e^{\beta E_{\pi}[u]+\beta \vec h\cdot\vec z} z_{i}\right|_{\vec h=0}.
\end{equation}
Therefore 
\begin{eqnarray}
\frac{1}{N}\sum_{i=1}^{N}\frac{\partial z_{i}}{\partial h_{i}} & = & \lim_{\beta\to\infty}\left.\frac{1}{\beta}\frac{\partial^{2}}{\partial h_{i}^{2}}\log Z(\beta,1,\vec h)\right|_{\vec h=0} \\
 & = & 
 \lim_{\beta\to\infty}\left.
 \frac{\beta}{N}\sum_{i=1}^{N}\left[E_{\beta,\vec h} \left[ z_{i}^{2}\right]-
 E_{\beta,\vec h} \left[ z_{i}\right]^{2}\right]\right|_{\vec h=0}\label{chidef11}
\end{eqnarray}
Here $E_{\beta,\vec h}[\ldots]$ stands for the expectation on the Gibbs measure $e^{\beta E_{\pi}[u]+\beta \vec h\cdot\vec z}$, and the second line arises from explicitly taking the derivative of $z_{i}$ as given in the second equality in Eq. (\ref{eqzi}). Comparing Eqs. (\ref{chidef1}) and (\ref{chidef11}) one concludes that Eq. (\ref{chidef}) holds true.
%\begin{equation}
%\label{chidef3}
%\chi=\frac{1}{N}\sum_{i=1}^{N}\frac{\partial z_{i}}{\partial h_{i}}
%\end{equation}
$\chi$ conveys information on the sensitivity of the solution to the specification of the problem parameters. A very similar argument can be made for the variation of $\vec z$ under a change $r_{i}^{\omega}\to r^{\omega}_{i}+\delta r_{i}$ with $h_{i}= E_{\pi}[u'(c^{\omega})/p^{\omega}]\delta r_{i}^{\omega}=\lambda \delta r_i$, which corresponds to a perturbation of the initial ($t=0$) prices of instruments.

\subsubsection{Wealth conservation and no-arbitrage condition}

Taking the expected value of Eq. (\ref{FOCc}), and using Eqs. (\ref{eqlam},\ref{eqkappa}), we find
\begin{equation}
\label{relaz}
1=\langle c^*p\rangle_{t,p}+\epsilon n\langle z^*\rangle_t
\end{equation}
As it should, the initial unit wealth goes partly to buy the consumption good and partly to banks, as a revenue for selling financial instruments.

If Eq. (\ref{FOCc}) is multiplied by $u'(c^{*})/p$, then, upon taking the expected value and using Eqs. (\ref{eqlam},...,\ref{eqkappa}), one can derive the relation
\begin{equation}
\label{noarb1}
\left\langle\frac{u'(c^*)}{p}(c^*p-1)\right\rangle_{t,p}=0.
\end{equation}
We recall that $u'(c^{\omega})/p^{\omega}$ is proportional to the equivalent martingale measure. Hence, this relation takes the form of a no-arbitrage condition: the value of the portfolio used to buy consumption at $t=1$ equals the initial wealth, i.e.$E_q[cp]=E_q[1]=1$. 

Relations (\ref{relaz}) and (\ref{noarb1}) constitute important consistency checks of the theory and show that important micro-economic constraints are built in the mathematical structure of the problem (\ref{maxUpsi}).

\subsubsection{The limit of complete markets}
\label{appA22}

The property $\langle f(t)t\rangle_{t}=\left\langle\frac{\partial f}{\partial t}\right\rangle_{t}$ of gaussian integrals, applied to Eqs. (\ref{eqnu},\ref{eqchi}), leads to
\begin{eqnarray}
\nu & = & \left\langle\frac{u"(c^*)}{\chi u"(c^*)-p^2}\right\rangle_{p,t}
\label{eqnu1}
\\
\chi & = & \frac{n}{\nu}{\rm Prob}\{z^*>0\}=\frac{\phi}{\nu}
\label{eqchi1}
\end{eqnarray}
we have used the expressions for $z^{*}$ and $c^{*}$ in Eqs. (\ref{FOCz},\ref{FOCc}). Hence,
\begin{equation}
\phi =  \left\langle\left[1-\frac{p^2}{\chi u"(c^*)}\right]^{-1}\right\rangle_{p,t}.
\label{eqnu2}
\end{equation}
Market becomes complete when $\phi=\nu\chi \to 1$. 
From Eq. (\ref{eqnu2}), it is clear that $\phi\to 1$ when $\chi\to\infty$. Leaving aside pathological cases (i.e. if $-\infty<u"(c)<0$), the converse is also true, $\chi\to\infty$ when $\phi\to 1$.

On the other hand, Eq. (\ref{eqchi1}) implies that
\[
\phi=\chi\nu=n\int_{t_0}^\infty\!\frac{dt}{\sqrt{2\pi}}e^{-t^2/2},~~~\hbox{with}~~~t_0=\frac{\epsilon\lambda}{\sigma}.
\]
With $\epsilon=0$ and $\sigma>0$, this implies $t_0=0$ and $\phi=n/2$. However, Eq. (\ref{eqnu2}) requires $\phi\le 1$, which means that $\lim_{\epsilon\to 0}\sigma>0$ can only hold for $n\le 2$. For $n>2$ the constraint $\phi\le 1$ requires that the extreme of integration $t_0$ in the above integral be finite and positive, i.e. that  $\sigma\to 0$ for $\epsilon\to 0$.

This implies that the line of complete markets is located at $\epsilon=0$ for $n\ge 2$, and that $\sigma=0$ on this line.

\subsection{The arbitrage region}
\label{appA3}

As discussed in appendix \ref{app0}, the volume $V$ of portfolios which violate the no-arbitrage condition, is given by
\[
\frac{1}{N}\log V=\lim_{\beta\to 0} \frac{1}{N}\log Z(\beta,x_0=0).
\]
So the evaluation of this quantity is directly obtained from the previous calculation.
Taking the average over the realization of the market,  we find
\begin{eqnarray}
h & \equiv & \lim_{N\to\infty}\frac{1}{N}\left\langle \log V\right\rangle\\
  & = & \max_{\omega,\chi,\lambda}\left[H_1(\omega,\chi,\lambda)+\max_{\sigma,\rho,\nu} H_2(\sigma,\rho,\nu|\omega,\chi,\lambda)\right]
\end{eqnarray}
where
\begin{eqnarray}
\label{bbb}
H_1&=&
  \frac{1}{n}\left\langle\log\left[\frac{1}{2}{\rm erfc}\left(\frac{\sqrt{n\omega}t+n\lambda}{\sqrt{2n\chi}}\right)\right]\right\rangle_t\\
  H_2&=& \frac{1}{2}\left[(\nu-\sigma^2-\rho^2)\chi+\nu\omega\right]+\rho\lambda+\left\langle\log\int_0^\infty\!dz e^{-\frac{\nu}{2}z^2+(\sigma t+\epsilon\rho)z}\right\rangle_t.
\end{eqnarray}
First notice that if $\vec z$ belongs to the volume, then also $a\vec z$ does. Indeed $h$ has 
the required invariance property for $z_i\to az_i$. Indeed, under the transformation $\chi\to a^2\chi$, $\omega\to a^2\omega$, $\lambda\to a\lambda$. $\rho\to\rho/a$, $\sigma\to\sigma/a$ and $\nu\to\nu/a^2$, both $H_1$ and $H_2$ are invariant.

The limit where the volume shrinks to zero is given by the case where the distance between two solutions goes to zero, i.e. $\chi\to 0$. In this limit, inspection of the saddle point equations reveals that $\omega,\lambda$ attain finite limits, whereas the other parameters behave as $\nu=v/\chi$, $\sigma=s/\chi$ and $\rho=r/\chi$. The leading behavior of $h$ must be of the type $\log\chi$, hence the terms proportional to $1/\chi$ must cancel. From this we derive the equation
\begin{eqnarray}
\tilde h & = & \lim_{\chi\to 0} h\chi =\tilde H_1+\tilde H_2\\
\tilde H_1 & = &  -\frac{\omega}{2n}\left\langle\theta(t+t_0)(t+t_0)^2\right\rangle_t,~~~~t_0=\sqrt{\frac{n}{\omega}}\lambda\\
\tilde H_2 & = & \frac{1}{2}[v\omega-s^2-r^2]+r\lambda+\left\langle\max_{z\ge 0}\left[-\frac{v}{2}z^2+(ts+r\epsilon)z\right]\right\rangle_t\\
& = & \frac{1}{2}[v\omega-s^2-r^2]+r\lambda-\frac{v}{2}
\left\langle {z^*}^2\right\rangle_t+s\left\langle t{z^*}\right\rangle_t+r\epsilon \left\langle {z^*}\right\rangle_t
\end{eqnarray}
The saddle points on $r,s$ and $v$ yield $\lambda=r-\epsilon\left\langle {z^*}\right\rangle_t$, $\omega=\left\langle {z^*}^2\right\rangle_t$ and $s=\left\langle {z^*}t\right\rangle_t$ which yield $\tilde H_2=\frac{s^2+r^2}{2}$ at the saddle point. In addition, $z^*=\frac{s}{v}(t+\xi)\theta(t+\xi)$, with $\xi=r\epsilon/s$. 
Expressing everything in terms of $\xi$, we find
\[
t_0(\xi)=\sqrt{\frac{n}{I_2(\xi)}}\left[\frac{\xi I_0(\xi)}{\epsilon}-\epsilon I_1(\xi)\right],~~~~I_n(\xi)=\left\langle \theta(t+\xi)(t+\xi)^n\right\rangle_t.
\]
and we can reduce the saddle point problem to one in the two variables $s$ and $t_0$
\[
\tilde h= \frac{s^2}{2}\left[1+\frac{\xi^2}{\epsilon^2}-\frac{I_2(\xi)}{nI_0(\xi)^2} I_2\left[t_0(\xi)\right]\right]
\]
notice that the saddle point equation in $s$ readily implies $\tilde h=0$ as assumed at the beginning.
The equation $\frac{\partial \tilde h}{\partial \xi}=0$ yields a second equation, which determines the relation between $n$ and $\epsilon$, on the boundary of the unstable region.

%
%Noticing also that $I_n'(x)=nI_{n-1}(x)$ for $n>0$, we can write
%\[
%\frac{d\tilde h}{d\xi}=\frac{2\xi}{\epsilon^2}-\frac{2I_2[t_0(\xi)]}{nI_0(\xi)^2}\left[I_1(\xi)-I_2(\xi)\frac{d\log I_0}{d\xi}\right]-\frac{2I_2(\xi)I_1[t_0(\xi)]}{nI_0(\xi)^2}\frac{dt_0}{d\xi}
%\]
%with 
%\[
%\frac{dt_0}{d\xi}=\sqrt{\frac{n}{I_2}}\left[\frac{I_0+\xi dI_0/d\xi}{\epsilon}-\epsilon I_0\right]-\frac{I_1}{I_2}t_0.
%\]
%I can solve trivially $\tilde h=0$ for $n$ and then I'm left with a single equation for $\xi$. No: t_0 depends on n

\subsection{Risk premium and competitive prices of assets}

The problem in Eq. (\ref{mineps}) can again be studied with the same techniques as above. In this case, one can build a partition function 
\[
Q(\beta)=\int \!d\vec w e^{-\beta H}
\]
with 
\[
H=\frac{\gamma}{2}\sum_{\omega=1}^\Omega\left(\zeta^\omega_{N+1}-\sum_{i=1}^K w_i \zeta_i^\omega\right)^2+\epsilon\sum_{i=1}^K w_i
\]
where $\zeta_i^\omega=r_i^\omega+\epsilon/\Omega$ are the fluctuations of asset returns. The sum on $i$ runs only on the traded assets (those with $z_i>0$). The number of these is given by $K=\phi\Omega$. 
The derivation below assumes that these can be considered as drawn independently, with the constraint $\sum_\omega \zeta_i^\omega=0$, and it neglects the possible bias due to the selection of consumers.
In the case $\sum_iw_i=0$, the residual risk is given by
\[
\min_{\vec w}\Sigma^2(\vec w)=-\frac{1}{\Omega}\lim_{\beta\to\infty} \log Q(\beta)
\]
In the limit of large $\Omega$ this limit satisfies a law of large numbers, i.e. it converges to its expectation on the realization of the variables $\zeta_i^\omega$. We skip the derivation, which follows exactly the same lines as in Ref. \cite{macroecondyn}, and state the result. The problem, in the case $\sum_iw_i=0$, reduces to the evaluation of the saddle point of the function
\[
\Gamma=\frac{\gamma(\Lambda^2-g-1)}{2(1+\gamma g)}+\lambda\Lambda+\frac{1}{2}g\nu+\frac{1}{2}\chi r+\frac{\phi}{2\nu}\left(\lambda^2-r\right)
\]
%\Gamma=\frac{\gamma(\Lambda^2-g-1)}{2(1+\gamma g)}+\lambda\Lambda+\frac{1}{2}g\nu+\frac{1}{2}\chi r+\frac{\phi}{2\nu}\left(\epsilon^2\Omega+\lambda^2-r)
Here we use the same symbols for variables which play an analogous role as in the other problems. 
The first order conditions, readily yield $\lambda=\Lambda=0$, $\nu=\gamma(1-\phi)$,
\[
g\equiv\sum_{i=1}^Kw_i^2=\frac{\phi}{1-\phi},~~~\hbox{and}~~~\chi=\frac{\phi}{\gamma(1-\phi)}.
\]
The value of the residual risk, when computed at the saddle point value, is given by $\Sigma^2=\frac{\gamma}{2\Omega}(1-\phi)$, which leads to Eq. (\ref{epssig})

\end{document}